\documentstyle[12pt]{article}
\advance\textheight by 90pt
\advance\voffset by -60pt
\advance\textwidth by 50pt
\advance\oddsidemargin by -25pt
\advance\evensidemargin by -25pt

\def\single_space{\baselineskip 12pt plus 1pt minus 1pt}
\def\one_and_a_half_space{\baselineskip 19pt plus 1pt minus 1pt}
\def\double_spacesp{\baselineskip 25pt plus 2pt minus 2pt}

\def\trade{{\bigcirc}\!\!\!\!\!\mbox{{\tiny R}}}
\def\mathmath{{\it Mathematica}$_{\trade}$\,\,}
\begin{document}
%\begin{titlepage}
\thispagestyle{empty}
\begin{flushright}
November 1999 \\
PSU/TH/221 
\end{flushright}
\vskip 1.cm
\double_spacesp

\begin{center}
{\Large {\bf
Comparing classical and quantum probability distributions for
an asymmetric infinite well }}
\end{center}
\begin{center}
M.~A.~Doncheski \footnote{mad10@psu.edu} \\
Department of Physics \\
The Pennsylvania State University\\
Mont Alto, PA 17237  USA \\
\end{center}
\begin{center}
and \\
R. W. Robinett \footnote{rick@phys.psu.edu}\\
Department of Physics\\
The Pennsylvania State University\\
University Park, PA 16802 USA \\
\end{center}
\vskip 0.5cm
\begin{abstract}
We compare the classical and quantum mechanical position-space 
probability  densities for a particle in an asymmetric infinite well.
In an idealized system with a discontinuous step in the middle of the
well, the classical and quantum probability distributions agree
fairly well, even for relatively small quantum numbers, 
 except for anomalous cases which are due to the unphysical
nature of the potential. We  are  able to derive upper and lower bounds on
 the differences  between the quantum and classical results.
We also qualitatively  discuss the momentum-space probability densities 
for this system using intuitive ideas about how much time a classical
particle spends in various parts of the well. This system provides an
excellent example of a non-trivial, but tractable,  quantum mechanical
bound state problem where the correlations between the 
amplitude and curvature 
of  quantum mechanical wavefunctions can be easily compared to classical
intuition about particle motion, with  quantitative success, but also warning 
of possible surprises in non-physical limiting cases.
\end{abstract}
%\end{titlepage}
\newpage

\setcounter{page}{1}

\double_spacesp

\begin{flushleft}
{\large {\bf 1.~Introduction}}
\end{flushleft}
\vskip 0.2cm

The calculation of position-space energy eigenstates in quantum mechanical
bound state systems is only the first step in student understanding of
the underlying physics. Very often,  a comparison of the resulting
quantum probability densities, $|\psi(x)|^2$ versus $x$, with classical
probability distributions  provides much needed insight into the deep and
pedagogically useful connections between the two approaches to mechanics. 
The most familiar
problem of one-dimensional quantum theory, namely the infinite well, 
is, for example,   occasionally
discussed in this manner \cite{liboff,robinett_book}. The local averaging
required to show that the oscillatory quantum probability 
distribution approaches the
well-known  flat classical value is easily implemented in this system, either
visually or more analytically, to illustrate in detail the correspondence
principle limit.

A far more familiar example and image in many modern physics and quantum
mechanics texts is the comparison of the classical and quantum results
for the harmonic oscillator eigenstates for increasingly large
quantum number, appearing in both older classic texts \cite{schiff,sherwin}
and as well as in very recent ones \cite{liboff,robinett_book}, 
\cite{serway_moses_moyer} -- \cite{cohen} 
at a wide variety of levels. In this case, both the amplitude and 
the curvature (`wiggliness') of the wavefunction vary in a non-trivial  
and highly correlated  way which can be understood using classical connections.
A number of  texts \cite{liboff, robinett_book},  \cite{beiser}, 
\cite{six_ideas} -- \cite{morrison}
now emphasize the intuitive ideas 
behind the form of wavefunctions as dictated by the shape  of the potential 
energy function and the value of the quantized energy eigenvalue, with
examples including less familiar systems such as linear potentials
\cite{robinett_book, goswami, morrison, robinett_1} and asymmetric
infinite wells \cite{robinett_book, krane, morrison}. Similar comparisons
are also possible for two-dimensional systems (such as the circular
infinite well \cite{robinett_2}) and an excellent discussion of the
 classical limit of the
quantum solutions for the hydrogenic radial probability distributions
has appeared in the pages of this journal \cite{rowe}. 

Most such presentations can easily give the impression that the
approach from the case of low-lying quantum states  states (small $n$) to the 
classical limit ($n >>1$)  is smooth and uninteresting and in this
note we wish to examine a simple system which exhibits some unexpected
properties. The potential we consider is an asymmetric infinite well defined 
via
\begin{equation}
      V(x)   = \left\{ \begin{array}{ll}
               \infty & \mbox{for $x<-a$} \\
   0 & \mbox{for $-a<x<0$} \\
   V_0 & \mbox{for $0<x<+b$} \\
               \infty  & \mbox{for $+b<x$}
  	                              \end{array}
\right.
\label{asymmetric_potential}
\end{equation}
so that it is an infinite well of width $(a+b)$, but with the right side
at a higher (constant) 
value of potential. (In most of our numerical calculations,
we will actually use values of $a=b$, but we will analyze the problem in some
generality,  at least initially.) Such a potential is shown in Fig.~1 
(with specific values of $a$, $b$, and $V_0$ for later use.)
This potential is useful for the
discussion of semiclassical limits for several reasons:

\newcounter{temp}
\begin{list}
{(\roman{temp})}{\usecounter{temp}}
\item It is a simple example where there is  a non-trivial 
but  easy-to-visualize variation of potential energy, and hence speed,
 between the classical turning points,
\item the classical concepts of how much time a particle spends
in each side of the well are intuitively obvious  so that the
classical probability distributions are straightforwardly obtained,
\item the quantum solutions can be obtained in closed form, and the resulting
energy eigenvalue conditions implemented numerically rather simply, and
\item the Fourier transform of the position space solutions can be readily
obtained to discuss the same qualitative and intuitive ideas in momentum
space.
\end{list}
It is not surprising, perhaps, that several textbooks 
\cite{tipler}, \cite{krane}, \cite{morrison} use this 
system as a qualitative example of intuitive wavefunction analysis.

\begin{flushleft}
{\large {\bf 2.~Quantum and Classical Solutions}}
\end{flushleft}
\vskip 0.2cm

To obtain the quantum solutions for this problem, we assume, at least
initially,   that $E> V_0$ (such as for those states 
labeled $5$ and above in Fig.~1), and we solve the time-independent 
Schr\"odinger equation in each side of the well, obtaining  solutions 
of the form
\begin{equation}
      \psi(x) = \left\{ \begin{array}{ll}
               A\sin[k(x+a)]  & \mbox{for $-a<x<0$} \\
               B\sin[q(x-b)]  & \mbox{for $0<x<+b$}
                                \end{array}
\right.
\label{general_solution}
\end{equation}
where
\begin{equation}
k = \sqrt{\frac{2mE}{\hbar^2}}
\qquad
\mbox{and}
\qquad
q = \sqrt{\frac{2m(E-V_0)}{\hbar^2}}.
\label{k_q_values}
\end{equation}
These solutions already satisfy the boundary conditions at the two infinite
walls. Insisting on the continuity of $\psi$ and $\psi'$ at the origin, 
the energy eigenvalue condition is given by 
\begin{equation}
k \cos(ka)\sin(qb) + q \cos(qb) \sin(ka) = 0
\label{eigenvalue_condition}
\end{equation}
which is easily solved graphically and/or numerically. In the case 
where $E<V_0$, we can let
\begin{equation}
q = \sqrt{\frac{2m(E -  V_0)}{\hbar^2}}
= i \sqrt{\frac{2m(V_0 - E)}{\hbar^2}}
= i\overline{q}
\end{equation}
and use the same solutions and eigenvalue condition, 
but  with the substitutions
\begin{eqnarray}
\sin(qb) & \rightarrow & \sin(i\overline{q}b) = i \sinh(\overline{q}b)
\nonumber  \\
\cos(qb) & \rightarrow & \cos(i\overline{q}b) =   \cosh(\overline{q}b)
\end{eqnarray}
and many multi-purpose mathematical packages such as \mathmath can
easily accommodate this change automatically. The energy eigenvalues can be 
generated for any given set of numerical parameters and the 
corresponding wavefunctions in (\ref{general_solution}) obtained
and normalized.

For purposes of comparison with purely classical results, we note that
the classical probability distribution for the particle when $E < V_0$
(where it would be restricted to bounce back and forth between
the walls at $-a$ and $0$) is
\begin{equation}
      P_{CL}(x)  = \left\{ \begin{array}{ll}
               1/a & \mbox{for $-a<x<0$} \\
               0 & \mbox{for $0<x<+b$}
                                \end{array}
\right..
\label{low_classical_probability}
\end{equation}
This implies that the probability of finding the particle in the left
side of the well in this case is 
\begin{equation}
\mbox{Prob}[-a<x<0] \equiv P_L^{(CL)} = 1
\qquad
\qquad
\mbox{for $E<V_0$}.
\label{left_probability_1}
\end{equation}

For the case when $E>V_0$, the situation is more interesting and we
can approach it by first calculating the time spent (classically) by
the particle in the left ($L$) and right ($R$) sides of the well, namely
\begin{eqnarray}
T_L = & 2a/v_L & = 2a\sqrt{\frac{m}{2E}} \nonumber \\
T_R = & 2b/v_R & = 2b\sqrt{\frac{m}{2(E-V_0)}}.
\end{eqnarray}
These combine to give the (classical) probability of finding the particle
in the left side as 
\begin{eqnarray}
\mbox{Prob}[-a<x<0] \equiv P_L^{(CL)}  = \frac{T_L}{T_L + T_R} 
   & = & \frac{a/\sqrt{E}}{a/\sqrt{E} + b/\sqrt{E-V_0}} 
\label{left_probability_2} \nonumber \\
& = & \frac{a\sqrt{E-V_0}}{a\sqrt{E - V_0} + b \sqrt{E}} \\ 
& = & \frac{a}{a + b\sqrt{E/(E-V_0)}} \nonumber
\end{eqnarray}
for $E>V_0$. (The final form is motivated by a discussion below.)
In a similar way we find that
\begin{equation}
\mbox{Prob}[0<x<+b] \equiv P_R^{(CL)} = 
\frac{b\sqrt{E}}{a\sqrt{E - V_0} + b \sqrt{E}}.
\end{equation}
In the limit where $E >> V_0$ (and the higher potential on one side has
little effect) we have the purely geometric results
\begin{equation}
P_{L}^{(CL)} \longrightarrow  \frac{a}{a+b}
\qquad
\mbox{and}
\qquad
P_{R}^{(CL)} \longrightarrow  \frac{b}{a+b}.
\end{equation}

Combining these results, we can evaluate the classical probability
distribution for the general $E> V_0$ case, namely
\begin{equation}
      P_{CL}(x)  = \left\{ \begin{array}{ll}
               \sqrt{E-V_0}/(a\sqrt{E-V_0} + b\sqrt{E})
 & \mbox{for $-a<x<0$} \\
               \sqrt{E}/(a\sqrt{E-V_0} + b\sqrt{E})           
     & \mbox{for $0<x<+b$}
                                \end{array}
\right.
\label{high_classical_probability}
\end{equation}
which is properly normalized since
\begin{equation}
\int_{-a}^{+b} \, P_{CL}(x)\,dx = 1.
\end{equation}
In order to compare these classical quantities to the quantum results,
we will eventually require the normalized, position-space energy
eigenstates to evaluate 
\begin{equation}
P_{L}^{(QM)} \equiv \int_{-a}^{0} \, |\psi(x)|^2\,dx
\qquad
\mbox{and}
\qquad
P_{R}^{(QM)} \equiv \int_{0}^{+b} \, |\psi(x)|^2\,dx.
\label{quantum_left_probability}
\end{equation}

As an example of the possible solutions in such a system, 
we first choose a standard set of parameters, namely
\begin{equation}
\hbar = 2m = 1
\qquad
\qquad
\mbox{and}
\qquad
\qquad
a = b = 3
\quad
,
\quad
V_0 = 20
\label{parameters}
\end{equation}
and examine the results in detail, indicating below how general they
are: specifically, we choose equal values of $a$ and $b$ to facilitate
comparison between the quantum and classical probabilities of finding
the particle in either side. With these parameters, the lowest $9$ energy 
eigenvalues are given by
\begin{equation}
E_n = 0.95\,,\, 3.78\,,\, 8.44\,,\, 14.78\,,\, 20.84\,,\, 22.34\,,\, 
24.94\,,\, 29.24\,,\, 33.30
\end{equation}
for $n = 1,...,9$ 
and these are the values shown in Fig.~1 (as horizontal dashed lines)
for the standard parameter set. (We note that the first four values,
corresponding to states mostly localized to the left side, have values
which are appropriately smaller than the first four states for an infinite
well of width  $a$, namely $E_n^{(\infty)}= \hbar^2 \pi^2 n^2/2ma^2$; 
with this parameter set, for example, we find that $E_n/E_n^{(\infty)}
\approx 0.85<1$ since the wavefunctions can penetrate into the
classically disallowed region, thereby reducing their `wiggliness' and
kinetic energy.)    We next plot, in Fig.~2, the
normalized position-space probability densities, $|\psi(x)|^2$
versus $x$, for these states, using the appropriately 
normalized eigenfunctions.
The vertical dashed line indicates the center of the well, while the
solid horizontal  lines (one only for the $E<V_0$ cases for $1-4$, 
and two for the $E>V_0$
cases for $5-9$) are the classical probability distributions in
(\ref{low_classical_probability}) and
(\ref{high_classical_probability}) respectively.

We first note the familiar general pattern of the addition of one `effective
half wavelength' or additional node as the quantum number increases,
typical of all such bound state systems. For the $1-4$ states,  which are
all below the $V_0$ threshold, the similarity to the corresponding
infinite well results is clear, with the additional feature of the
increased tunneling into the classically disallowed region. The first state
above threshold, $n=5$, is consistent with the usual intuitive rules about
the behavior of wavefunctions in different potential regions. The 
wavefunction is wigglier (less wiggly) but with lower (higher) amplitude
in the left (right) side where the classical kinetic energy and speed
is larger (smaller) and the classical particle would expected to
spend less (more) time. 
 The $n=5$ state is even roughly consistent with the
classical result (\ref{high_classical_probability}),
as far as such a low-lying state can be. Similar
expected patterns (and qualitative  agreement) are evident in the $n=7$ 
and $n=8$ states.

However, we note that the $n=6$ wavefunction has an unexpected form, 
at least in terms of the relative values of the amplitudes 
 on the left and right sides. In this case (as well as that for $n=9$ where 
a similar effect is seen to occur)
the wavefunctions being matched at the (discontinuous) $x=0$
boundary are connected (smoothly, of course) at what happens to be
very close to an antinode. Since the amplitude in each side of the well is
constant (since the potential is piecewise constant), if the wavefunctions
match at an anti-node, the amplitudes must match ($A=B$ in 
(\ref{general_solution})) everywhere in the well. 
Thus, while the `wiggliness' of the
wavefunction agrees with naive expectations based on considerations of
classical speed and/or kinetic energy, the amplitudes and resulting
probability densities are somewhat anomalous in that the particle will
be found much more frequently (about half of the time in this case) in
the left side of the well than expected from purely classical  arguments.
(Presumably, a student asked to sketch a possible solution in this
potential well, as in Refs.~\cite{tipler,krane}, and who provided
something like this $n=6$ case would not have received full credit!)

To discuss  these effects more quantitatively, we calculate the quantum 
probability
of finding the particle in the left-half of the well 
(\ref{quantum_left_probability}) for a number of states in this
model system for comparison with the classical results given
by (\ref{left_probability_1}) and  (\ref{left_probability_2}). We plot,
in Fig.~3, the quantum values (diamonds), $P_{L}^{(QM)}$, 
for all states with $E<100$
(up to $n=18$ in this case) for comparison to the solid curves
representing the classical result. The horizontal dashed line indicates
the high energy limit of $P_L = 0.5$ (when $a=b$) 
where the presence of a small
potential `bump' at the bottom of a very deep well (when $E>>V_0$) will
have little effect. We first note that for the states below threshold,
the quantum probabilities,  $P_L^{(QM)}$, decrease from near the purely 
classical value of $P_L^{(CL)} =1$ due to the increasing amount of quantum 
tunneling evident in Fig.~1. Above threshold, the quantum probabilities seem 
to  track the classical prediction
(\ref{left_probability_2}), except most dramatically for the
anomalous cases like $n=6,9$. In these instances, $P_L^{(QM)}$ is much 
larger than expected classically, but never exceeds $P_L=0.5$ which would truly
be unphysical as it would imply (classically) that the particle is moving 
slower in the left side of the well and therefore spending more time there.
A similar, but somewhat less dramatic, effect is also evident for those
states with quantum probabilities which are significantly less than the 
classical predictions, such as the $n=8$ state in Fig.~3. In all cases
we have studied where $P_L$ is most obviously much smaller  than
classical expectations, we have found  that the wavefunction matching occurs 
very close to a node. The two extreme cases of anomalously large
(small) values of $P_L^{(!M)}$ have thus been found to be connected to 
wavefunction matching near antinodes (nodes) at the discontinuous
boundary. 

Using this observation, we have been able to derive upper and
lower bounds for $P_L^{(QM)}$ which we have found to be satisfied
in all of our numerical studies. 
For example, in the case of matching at an antinode,
the boundary condition on the left/right amplitudes (from the continuity of
$\psi$) is $A=\pm B$ and the
resulting quantum probabilities can be evaluated to find that
\begin{equation}
P_{L}^{(QM,max)} = \frac{a}{a+b}
\label{upper_bound}
\end{equation}
which is also the high energy $(E>>V_0$) limit. For the case of
wavefunction matching at nodes, the appropriate condition comes from
the continuity of $\psi'$, namely $kA = \pm qB$, which then gives
\begin{equation}
P_{L}^{(QM,min)} = \frac{a}{a + b[E/(E-V_0)]}.
\label{lower_bound}
\end{equation}
(These results are considerably more simple in form than the
general expression for $P_{L}^{(QM)}$ since we are integrating over 
integral numbers of half- and quarter-wavelengths in each
side of the well in these special cases.)
These two forms can be compared to the purely classical result
(\ref{left_probability_2}) and we note that
\begin{equation}
\mbox{(node case)}
\qquad
P_{L}^{(QM,min)} \leq  P_{L}^{(CL)} \leq P_{L}^{(QM,max)} 
\qquad
\mbox{(antinode case)}
\end{equation}
since $\sqrt{E/(E-V_0)} > 1$. 
To illustrate these bounds, we have plotted $P_L^{(QM,max)}$
(horizontal dashed line) and $P_{L}^{(QM,min)}$ (dotted curve) in
Fig.~3 where they clearly  bracket the classical result. We find that 
the results presented there are very typical of all the cases we have 
studied, namely that the quantum results for $P_{L}^{(QM)}$ rigorously 
fall between the upper and lower bounds, often coming very close to 
saturating them.

\begin{flushleft}
{\large {\bf 3.~Quantum Results in a Smoothed Asymmetric Infinite Well}}
\end{flushleft}
\vskip 0.2cm

An obvious question about these effects (especially the anomalously
large values of $P_{L}^{(QM)}$) is whether they are  artifacts
of the specific set of model parameters (\ref{parameters}) used here
or a more general phenomenon related to the structure of the potential
well itself.  We have
repeated our analysis for a large number of different values of
$a,b$, and $V_0$ (including many cases where $a\neq b$) and have almost always
found that in the first 10-15  eigenstates above 
the $V_0$ threshold that there are 1-2 solutions which exhibit the 
wavefunction matching near  antinodes which lead to
anomalously large values of $P_L^{(QM)}$, compared to the
classical result (\ref{left_probability_2}), almost saturating
the upper bound (\ref{upper_bound}), as well as 1-2 states which come
close to saturating the lower bound (\ref{lower_bound}).
 We think that it is easy
to argue that these  effects  are  a result of the discontinuous (and hence
unphysical) nature of the potential step at $x=0$. In order to test this,
we have solved the Schr\"odinger equation for a `smoothed' version of
this asymmetric well given by
\begin{equation}
      V_1(x)  = \left\{ \begin{array}{ll}
               \infty & \mbox{for $x<-a$} \\
V_0/(1+e^{-x/\delta})   & \mbox{for $-a<x<+b$} \\
               \infty & \mbox{for $+b<x$}
                                \end{array}
\right..
\label{smooth_potential}
\end{equation}
This version still has impenetrable walls at $x=-a,+b$, but 
gives a smoother transition between $V(-a) \approx 0$ and
$V(+b) \approx V_0$ depending on the value of $\delta$: the discontinuous
step potential is recovered in the limit that $\delta \rightarrow 0$. 
This smoothed version is shown in Fig.~4 (middle) for 
$\delta = 0.2$ (as the dashed curve) along with the discontinuous 
($\delta = 0.0$) step potential. We then solve the Schr\"odinger
equation (numerically) to find the allowed energy eigenvalues and the
resulting normalized eigenstates for the smoother case.
 The results for the $n=6$ (top)
and $n=7$ (bottom) states are also shown in Fig.~4 where the
wavefunctions for the $\delta = 0.2$ cases are shown (dashed curves) 
for comparison to the discontinuous case (solid curves.) For the anomalous
$n=6$  case, even the introduction of a small bit of `smoothness' into the
potential allows the wavefunction to accommodate to the required matching 
and the resulting quantum probability density is much closer to the
classical expectation. On the other hand, for the $n=7$ case,  which was 
already in fairly good agreement with the classical result, the 
changes are much less dramatic. Similar smoothing functions yield
similar results as we can see by considering  a linear smoothing potential 
given by 
\begin{equation}
      V_2(x)  = \left\{ \begin{array}{ll}
               0 & \mbox{for $-a<x<-\epsilon$} \\
               V_0(1+x/\epsilon)/2 & \mbox{for $-\epsilon<x<+\epsilon$} \\
               V_0 & \mbox{for $+\epsilon<x<+b$}
                                \end{array}
\right..
\label{smooth_potential_2}
\end{equation}
If we expand both potentials (\ref{smooth_potential}) and
(\ref{smooth_potential_2})  near the origin ($x \approx 0$) we find that
they have a similar form, namely
\begin{equation}
V_1(x) \approx \frac{V_0}{2}\left(1 + \frac{x}{2\delta}\right)
\, \longleftrightarrow \, 
\frac{V_0}{2}\left(1 + \frac{x}{\epsilon}\right)
\approx V_2(x).
\label{potentials_agree}
\end{equation}
For numerical purposes, therefore, we will use $\epsilon = 2\delta$
for comparisons between the two smoothings. For example, in Fig.~4, we
include the linear extrapolation (center) and the results for the
$n=6$ (top) and $n=7$ (bottom) states as dotted curves (for
$\epsilon = 0.4$). Not surprisingly,
perhaps, the results are similar to the exponential smoothing.

In order to quantify the improvement 
in agreement with the classical results, we show in Fig.~5 the
quantum probabilities,  $P_L^{(QM)}$, for both the discontinuous
($\delta = \epsilon = 0.0$, diamond data as before) case
and a smooth exponential ($\delta = 0.2$, starred data)
case, compared again to the classical result (\ref{left_probability_2}).
The agreement, in general, is much better for the more physical potential
gradient, except very near threshold (the $n=5$ case) where small changes
in the shape of the potential might be expected to have large results.
We note that in each case (except for $n=5$) that there is 
improvement (often substantial) in the agreement with the classical result. 

At the same time, we show in Fig.~5 (top) the fractional
change in the energy eigenvalue for the new smoothed potential
compared to the discontinuous case, namely $\Delta E/E$, and note that
while the energies are different by no more than $3\%$ in the smoothed
potential, the changes in the corresponding values of $P_L^{(QM)}$ are
much more substantial, up to $100\%$. This implies that the introduction
of the smooth (and hence more physical) potential step 
 does not change the intrinsic properties of the system
 (energies, etc.) very much, 
but does allow for a more realistic description of the quantum probability 
densities and their approach to the classical limit. We find that in almost
all cases where this type of anomalous behavior is encountered
that the introduction
of smoothing with a length scale $\delta \sim \epsilon \sim \lambda/4$
(where $\lambda$ is the wavelength in the left side of the well)
is enough to substantially improve agreement with the classical probability 
predictions while making little change in the energy eigenvalues.

\begin{flushleft}
{\large {\bf 4.~Momentum Space Results}}
\end{flushleft}
\vskip 0.2cm

We can also develop our semi-classical intuition, as well as observing
some of the anomalous quantum behavior, by examining the 
momentum-space probability densities for this system. The momentum-space
wavefunctions, $\phi(p)$, can be obtained from the $\psi(x)$ 
in (\ref{general_solution}) using  the Fourier transform via
\begin{equation}
\phi(p) = \frac{1}{\sqrt{2 \pi \hbar}} 
\int_{-\infty}^{+\infty} \psi(x) \, e^{-ipx/\hbar} \, dx.
\end{equation}
For the $\delta =0.0$ case,  where we can use the analytic results 
(\ref{general_solution}), the resulting normalized momentum-space
probability distributions are easy to generate and we show them for the
same first $9$ eigenvalues in Fig.~6. The vertical dashed lines 
correspond to  the values of $p = \pm \hbar k$ (for all values of $E$) 
while the dotted lines indicate values of $p = \pm \hbar q$
(for states with $E>V_0$) where we would expect to see features corresponding
to classical, back-and-forth  motions in the left side ($p = \pm \hbar k$) and 
right sides ($p = \pm \hbar q$) of the well. For the states below threshold,
the results are similar to familiar ones
\cite{robinett_book,robinett_1} for the standard infinite well. For the
$n=5$ case, the first above threshold, 
 we see a central feature, much like that for $n=1$, as well
as two small features at larger  values of $|p|$. The similarity between
the $E_1 = 0.95, k = 0.974$ central feature for $n=1$ and the
$E_5-V_0 = 0.84, q = 0.914$ central peak for $n=5$ is clear. The smaller
features at larger values of $p = \pm \hbar k$ are qualitatively consistent
with the small  amount of time spent in the left-side of the well for
this case where $q<<k$.  The anomalously large value of $P_L^{(QM)}
\approx 0.5$
for the $n=6$ case, where the particle spends  roughly equal amounts of time
in each side of the well, is evident in the approximately equal magnitudes
of the $q,k$ `bumps' in this case. We note that in general we expect to
see slightly larger (smaller)  features at the smaller (larger)
values of $p = \pm \hbar q$ ($p = \pm \hbar k$) due to our intuitive
probability arguments ($P_R >  P_L$) and this pattern is also reasonably 
evident for the $n=7,8$ cases. This pattern becomes increasingly difficult to
identify, however,  as $n$ increases because the $k,q$ features tend to merge
due to the fact that
\begin{eqnarray}
k - q \propto \sqrt{E} - \sqrt{E-V_0} 
& = & \sqrt{E} \left(1 - \sqrt{1-\frac{V_0}{E}}\right) \nonumber \\
& = & \sqrt{E} \left( 1 - 1 + \frac{V_0}{2E} + \cdots \right) \\
& \approx & \frac{V_0}{2\sqrt{E}} 
\nonumber 
\end{eqnarray}
which goes to zero as $E$ increases.
The difficulty with unambiguously identifying them is already evident in 
the $n=9$ case where the classical $k,q$ features are obvious, but the 
interference between the $k,q$ terms gives rise to a feature between
them which is even larger.

It is tempting to imagine trying to correlate, in a quantitative manner,
the `amount of probability' in the $\pm q$ and $\pm k$ `peaks' with the
position-space probabilities of measuring the particles in the
right and left sides of the well respectively. We find, however, that 
given the large amount of interference between the $k,q$ pieces of the
Fourier transform necessary to obtain $\phi(p)$ that no such identification
is possible.

\begin{flushleft}
{\large {\bf 5.~Conclusions}}
\end{flushleft}
\vskip 0.2cm

In conclusion, we have identified a simple model quantum mechanical
bound state system, the asymmetric infinite well, which can be easily analyzed
and for which the comparison of the quantum and classical probability
distributions finds many points of similarity, even exhibiting
quantitative  agreement for probabilities for many states. We have even
been able to derive seemingly rigorous upper and lower bounds on how much
the quantum results deviate from the classical probability expectations.
The system can also be used to qualitatively examine classical intuition about
momentum--space probabilities as well, given the relative simplicity 
of the position-space wavefunctions. However, it seems to yield, in almost 
every case studied, some solutions which have
anomalous probability distributions, compared to classical expectations,
due to the unphysical (discontinuous) nature of the potential. It also serves,
therefore, as something of a cautionary tale about the use of idealized
models.

\begin{flushleft}
{\large {\bf Acknowledgments}}
\end{flushleft}
\vskip 0.2cm

The work of MAD was supported, in part, by the Commonwealth College of The
Pennsylvania State University under a Research Development Grant (RDG); 
the work of RR was supported, in part, by NSF grant DUE-9950702.

\newpage

\begin{flushleft}
{\Large {\bf 
Figure Captions}}
\end{flushleft}
\vskip 0.5cm
 
\begin{itemize}
\item[Fig.\thinspace 1.]  Example of the asymmetric infinite well potential
(\ref{asymmetric_potential})
for the special case of $a=b=3$ and $V_{0} = 20$. The nine lowest energy
levels, obtained from the energy eigenvalue
equation (\ref{eigenvalue_condition}) with the parameters in
(\ref{parameters}),  are shown as horizontal dashed lines.
\item[Fig.\thinspace 2.] Normalized position-space probability densities,
$|\psi(x)|^2$ versus $x$, for the first nine energy levels in the
asymmetric well shown in Fig.~1 with the parameters in 
(\ref{parameters}).  The solid horizontal lines indicate
the classical probability distributions 
 (\ref{low_classical_probability}) (for $E<V_0$ and states $1-4$) and
(\ref{high_classical_probability}) (for $E>V_0$ and states $5-9$). 
The vertical dashed lines indicate  the center of the well at $x=0$.
\item[Fig.\thinspace 3.] Classical probability of finding the particle 
in the left side of the well (solid line and curve), given by 
(\ref{left_probability_1}) (for $E<V_0$ and states $1-4$) and 
(\ref{left_probability_2}) (for $E>V_0$ and states $5-9$),
as a function of $E$ for the states in Figs.~1 and 2 with the
parameters in (\ref{parameters}). The data points (diamonds) are
the corresponding quantum probabilities,  $P_{L}^{(QM)}$ 
(\ref{quantum_left_probability}),  for the lowest $18$ energy eigenvalues
(all those with $E<100$) with the standard parameter set. The horizontal
dashed line at $P_L = 0.5$ corresponds to equal amounts of time spent
in the left and right halves of the well which is also the upper bound
(\ref{upper_bound}),  $P_{L}^{(QM,max)}$, for this case. The dotted curve
is the corresponding lower bound (\ref{lower_bound}),
$P_{L}^{(QM,min)}$.

\item[Fig.\thinspace 4.] The middle figure 
shows two  `smoothed' versions  of the discontinuous step (solid curve)
at the center of the well given
by (\ref{smooth_potential}) ($\delta = 0.2$, dashed curve) and
(\ref{smooth_potential_2}) ($\epsilon = 0.4$, dotted lines).
The value  of $\epsilon = 2\delta$ is chosen so that the
two smoothings agree near $x\approx 0$ as in 
(\ref{potentials_agree}). The solutions corresponding to 
$n=6$ (top) and $n=7$ (bottom) for the discontinuous well 
($\delta = \epsilon = 0.0$, solid curves) 
and the smooth cases  with $\delta = 0.2$ (dashed curves)
in (\ref{smooth_potential}) and  $\epsilon = 0.4$
(dotted curves) in (\ref{smooth_potential_2}) are shown. 
The `anomalous' $n=6$ case is affected dramatically and
becomes much more consistent with the classical predictions for the
smoothed cases, while the more standard $n=7$ case is not affected very much.
\item[Fig.\thinspace 5.] Same as Fig.~3, but for $E>V_0$ only. The
solid curve is the classical prediction for $P_L^{(CL)}$, and 
the results for the quantum probability from
 (\ref{quantum_left_probability}) for the discontinuous case
($\delta = 0.0$, diamonds again) and the smoothed case 
(\ref{smooth_potential}) with
$\delta = 0.2$ (stars) are shown. The upper (dashed) and lower
(dotted) bounds on $P_{L}^{(QM)}$ from 
(\ref{upper_bound}) and (\ref{lower_bound}) are indicated.
Even a small amount of smoothing
of the discontinuity improves the agreement with the expected
classical result dramatically. The fractional change  in energy
$\Delta E/E$ induced for each state by using $\delta = 0.2$ compared
to $\delta = 0.0$ is shown at the top. In no case is the change larger
than $3\%$ and it decreases quickly as the energy is increased.
\item[Fig.\thinspace 6.] Normalized momentum-space probability
distributions, $|\phi(p)|^2$ versus $p$, for the lowest nine states
in the asymmetric well with the standard parameter set 
(\ref{parameters}). The vertical dashed lines show values of
$p = \pm \hbar k$ (for all $E$) and the
dotted lines correspond to  $p = \pm \hbar q$ (for $E>V_0$) from 
(\ref{k_q_values})  for comparison. The
momentum-space wavefunctions are calculated by taking the Fourier
transform of the position-space solutions in (\ref{general_solution}).
\end{itemize}

\newpage

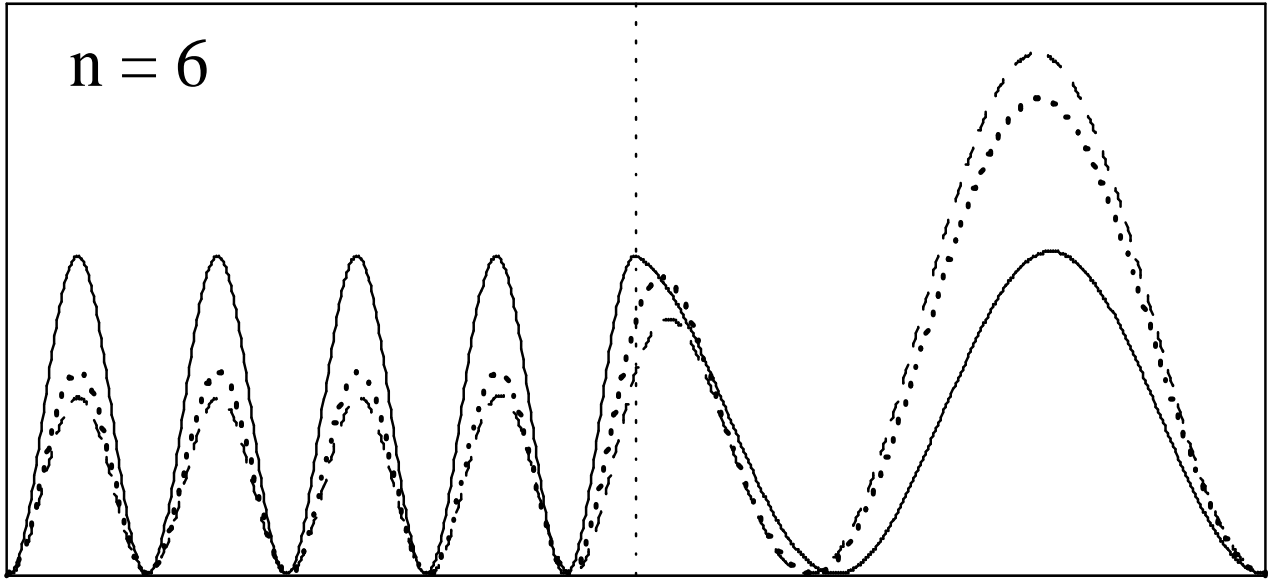
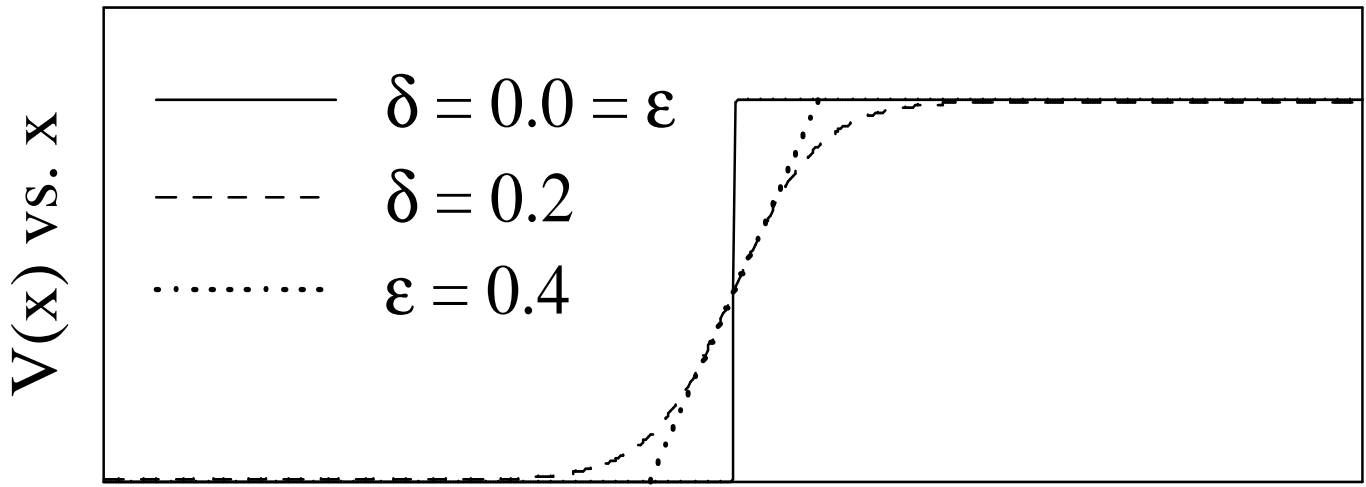
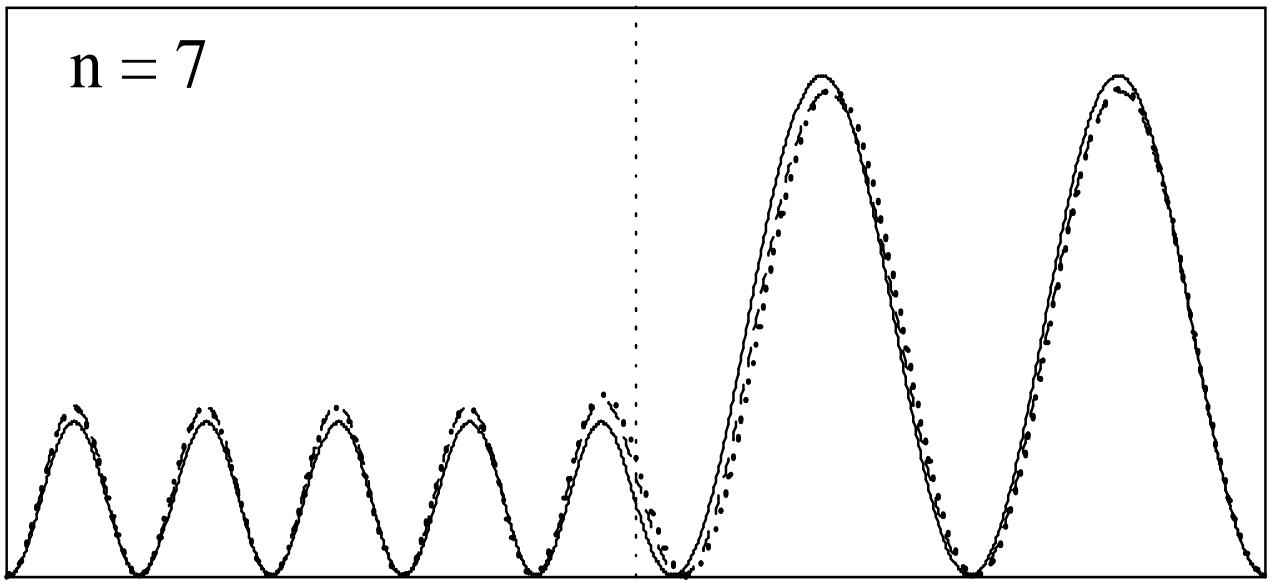
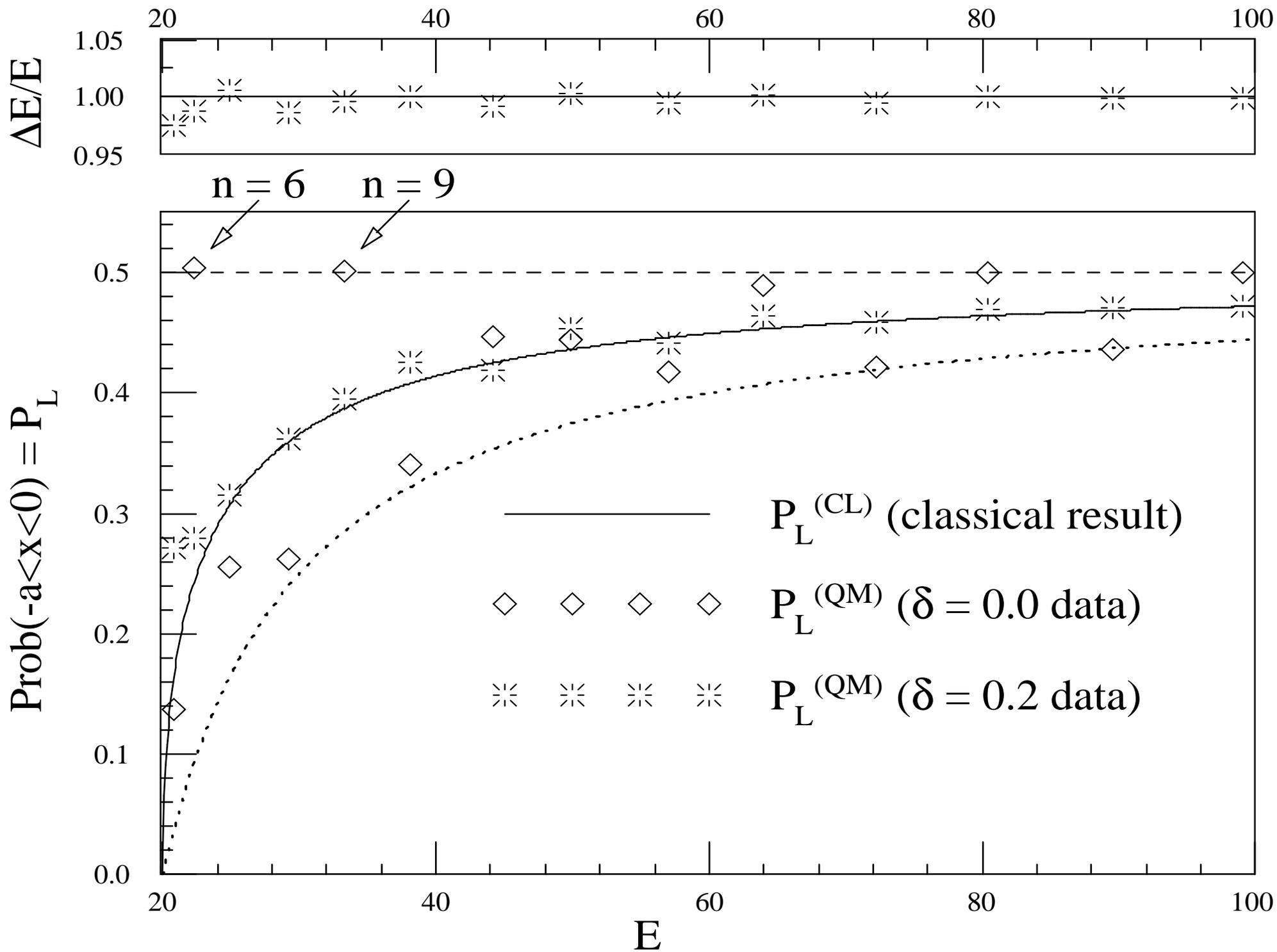
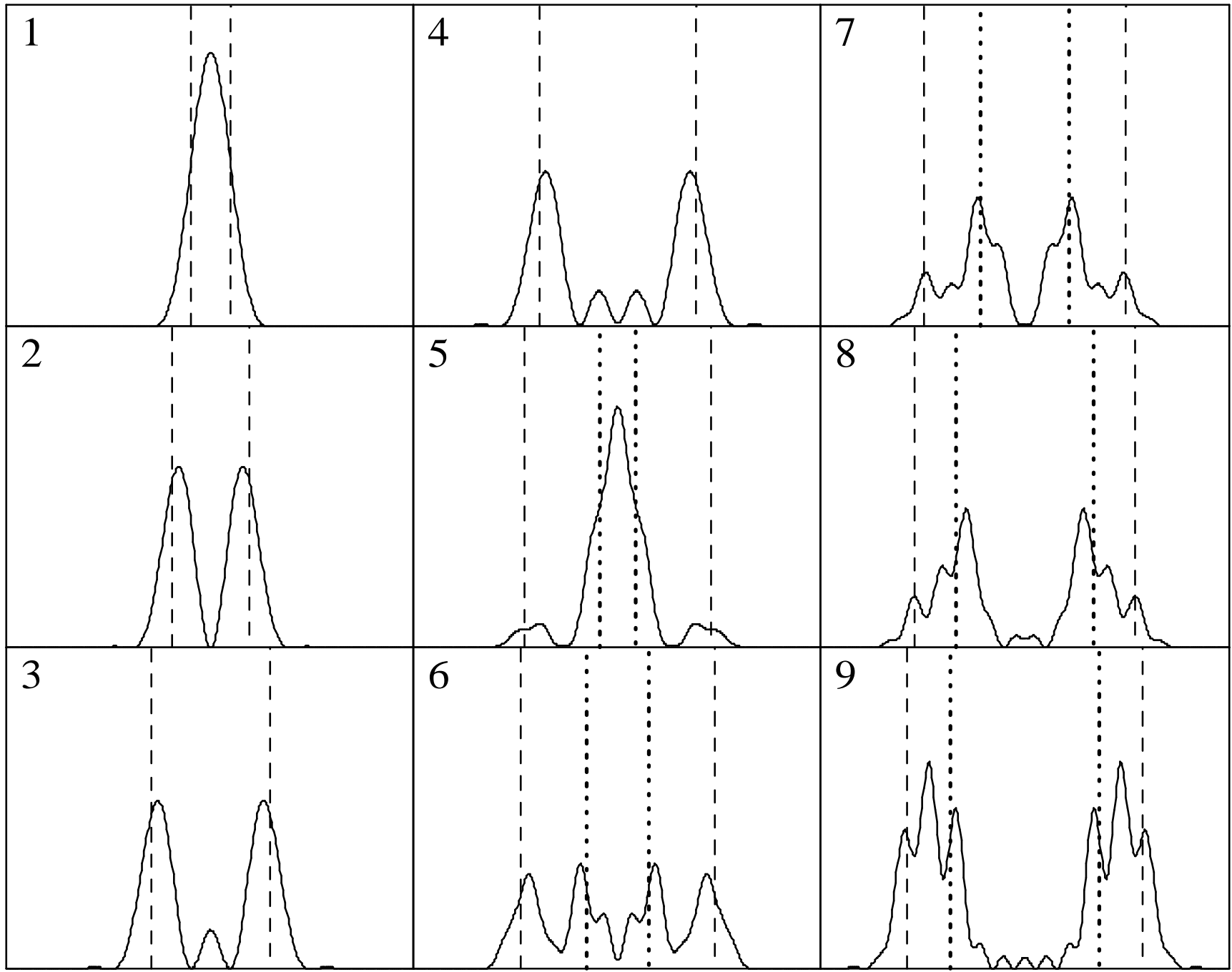
\end{document}